\documentclass[%
 reprint,superscriptaddress,showpacs,preprintnumbers,amsmath,amssymb,aps,pra,showkeys]{revtex4-1}
\usepackage{graphicx}
\usepackage{amsmath}

\usepackage{graphicx}
\usepackage{dcolumn}
\usepackage{bm}
\usepackage{graphics}
\usepackage[final]{hyperref}
\usepackage{amssymb}
\usepackage{textcomp}
\usepackage[floats=float]{chemscheme}

\begin{document}
\preprint{published in: Chem. Commun. 46, 4145-4147 (2010). \href{http://dx.doi.org/10.1039/c0cc00125b}{DOI: 10.1039/c0cc00125b}}
\title{Quantum interference distinguishes between constitutional isomers}

\author{Jens T\"uxen}
\affiliation{University of Basel, Department of Chemistry, St. Johannsring 19, 4056 Basel, Switzerland}
\author{Stefan Gerlich}
\author{Sandra Eibenberger}
\author{Markus Arndt}%
\homepage{http://www.quantumnano.at} \email{markus.arndt@univie.ac.at} \affiliation{University of Vienna, Faculty of Physics, Boltzmanngasse 5,
1090 Vienna, Austria}
\author{Marcel Mayor}
\email{marcel.mayor@unibas.ch}\affiliation{University of Basel, Department of Chemistry, St. Johannsring 19, 4056 Basel, Switzerland}
\affiliation{KIT Karlsruhe, Institute for Nanotechnology, P. O. Box 3640, D-76021 Karlsruhe}

\begin{abstract}
Matter waves, as introduced by de Broglie in 1923~\,\cite{Broglie1923a}, are a fundamental quantum phenomenon, describing the delocalized center of mass motion of massive bodies and we show here their sensitivity to the molecular structure of constitutional isomers.
\end{abstract}


\maketitle

In quantum textbooks, matter wave phenomena are often associated with the equation $\lambda_{dB}=h/(mv)$, where the de Broglie wavelength $\lambda_{dB}$ is only determined by Planck's constant $h$ and the particle's momentum $p=mv$. It is a common conjecture that this relation still holds for bodies of arbitrary mass, size and complexity. But what is the role of the detailed internal molecular composition if $\lambda_{dB}$ does not include any such information?

As already shown before~\,\cite{Hackermueller2004}, the centre of mass motion can be well described by quantum delocalization and interference of each entire molecule with itself --- even if the internal atoms may populate a thermal mixture of vibrational and rotational modes at a temperature of about $T=500\,$K. The conceptual separation of the internal and external degrees of freedom allows us to discuss two different cases:

If the interaction between the molecule and its environment makes it possible, even only in principle, to retrieve position information, the initially delocalized particle will be localized and quantum interference can no longer occur~\,\cite{Chapman1995}.

The second case is of particular relevance for our present experiment: All individual atoms in a given molecular structure will add up to determine its global properties, and in particular also its electrical susceptibility. The susceptibility can couple to an external electric field and thus influence the center of mass motion of the entire particle. In this way, the internal structure becomes influential for the external motion, even though the molecule remains still widely delocalized and capable of showing de Broglie interference.

First experiments along this line were recently able to apply this idea to near-field interferometry for measuring the static~\,\cite{Berninger2007} and optical~\,\cite{Hackermueller2007} polarizability $\alpha_{\mathrm{stat}}$ and $\alpha_{\mathrm{opt}}$ as well as the total electric susceptibility $\chi_{\mathrm{tot}}$ of molecules. The latter may also contain additional information about static or time varying electric dipole moments~\,\cite{Compagnon2002}. It has therefore been suggested that different molecular conformations might eventually also be distinguished in quantum interference experiments~\,\cite{Ulbricht2008a}.

For demonstration purposes we here compare two tailor-made model compounds \textbf{1} and \textbf{2}. The design of both is based on our recent findings in molecular electronics which showed a considerable delocalization of the $\mathrm{\pi}$-system in rod-like oligo phenylene ethynylene (OPE)~\,\cite{Wu2008a,Huber2008} on the one hand and a partition of the $\mathrm{\pi}$-system by conjugation interrupting subunits like e.g. platinum complexes~\,\cite{Mayor2002}, perpendicular torsion angles~\,\cite{Vonlanthen2009,Loertscher2008} or sp$^3$-carbon atoms on the other hand. The constitutional isomers \textbf{1} and \textbf{2} both have the chemical formula C$_{49}$H$_{16}$F$_{52}$ and an identical molecular weight of $1592\,\mathrm{g\,mol^{-1}}$, but different polarizabilities. The tetrahedral model compound \textbf{1} consists of four phenyl rings interlinked by a central sp$^3$-carbon which disables conjugation between the neighboring aromatic subunits. In contrast to that, the $\mathrm{\pi}$-system of the rod-like OPE \textbf{2} extends over all three phenyl rings. Thus an increased electron delocalization and polarizability is expected for compound \textbf{2} as compared to \textbf{1}.

A simulation using Gaussian 03W~\,\cite{Frisch2003short} with the basis set 3-21G yielded a static polarizability of $\alpha_{\mathrm{stat}}= 63(2)\,\mathrm{\AA^3\times4\pi\epsilon_0}$ for compound \textbf{1} and $70(2)\,\mathrm{\AA^3\times4\pi\epsilon_0}$ for molecule \textbf{2}, i.e. a difference of about $10\,$\% between the two structures.

The tetrahedral target compound \textbf{1} was synthesized in two steps starting from tetraphenylmethane (\textbf{3}) (Scheme \ref{fig:Scheme1}). Bromination of all four \textit{para}-positions of tetraphenylmethane (\textbf{3}) was achieved according to an established procedure~\,\cite{Rathore2004}. Pure tetrakis(4-bromophenyl)methane (\textbf{4}) was isolated as a yellowish crystalline solid in a yield of 83\%. By treating \textbf{4} with \textit{n}-perfluorohexyl iodide in a suspension of copper in dimethylformamide (DMF) at $120\,$\textdegree C the perfluoroalkyl chains were introduced. The tetrahedral target structure tetrakis(4-perfluorohexylphenyl)methane (\textbf{1}) was isolated by column chromatography (CC) as a white crystalline solid in $27\,$\% yield.	

\begin{scheme}[b]
  \begin{center}
\includegraphics*[width=1.0\columnwidth]{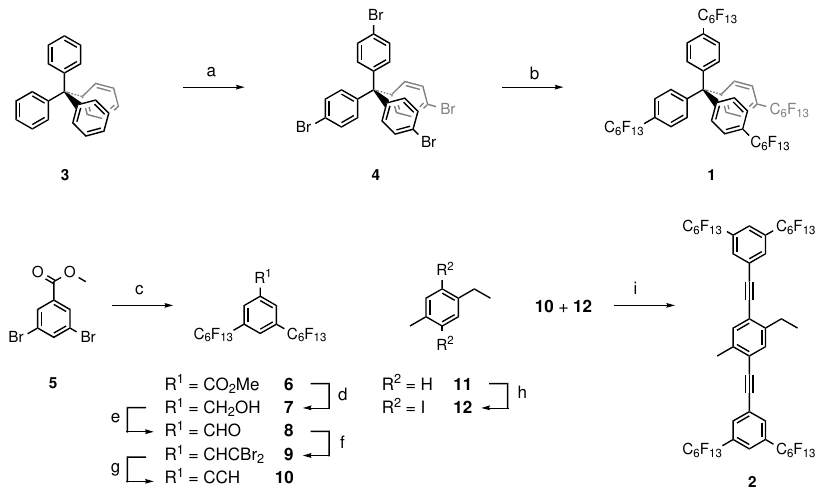}
  \end{center}
  \caption{Synthesis of the structural isomers \textbf{1} and \textbf{2}. (a) Br$_2$, rt, $25\,$min, $83\,$\%; (b) IC$_6$F$_{13}$, Cu, DMF, $120\,$\textdegree C, $12\,$h, $27\,$\%; (c) IC$_6$F$_{13}$, Cu, DMF, $120\,$\textdegree C, $12\,$h, $86\,$\%; (d) LiAlH$_4$, Et$_2$O, rt, $16\,$h, $64\,$\%; (e) PDC, CH$_2$Cl$_2$, rt, $24\,$h, $88\,$\%; (f) CBr$_4$, PPh$_3$, CH$_2$Cl$_2$, $0\,$\textdegree C, $1\,$h, $85\,$\%; (g) 1. \textit{n}-BuLi, THF, $-78\,$\textdegree C; 2. H$_2$O, $93\,$\%; (h) I$_2$, HIO$_3$, glacial AcOH, H$_2$SO$_4$, CHCl$_3$, H$_2$O, $85\,$\textdegree C, $4\,$h, $53\,$\%; (i) Pd(PPh$_3$)$_4$, CuI, (\textit{i}-Pr)$_2$NH, THF, rt, $12\,$h, $22\,$\%.}
\label{fig:Scheme1}
\end{scheme}

The rod-like OPE structure \textbf{2} was assembled starting from methyl 3,5-dibromobenzoate (\textbf{5}). In a copper catalyzed Ullmann type coupling reaction both bromines were substituted with fluorous ponytails to provide \textbf{6} as a white solid in $86\,$\% yield after CC. The benzylic alcohol \textbf{7} was obtained in $64\,$\% yield by reduction with lithium aluminium hydride (LiAlH$_4$). Subsequent oxidation with pyridinium dichromate (PDC) provided the benzaldehyde \textbf{8} as a white solid in $88\,$\% yield after CC. With a Corey-Fuchs reaction sequence the aldehyde was transformed to the alkyne \textbf{10} in $79\,$\% yield over both steps. The diiodo derivative \textbf{12} as central building block of the rod \textbf{2} was obtained by iodination of 4-ethyltoluene (\textbf{11}) as white crystals in $53\,$\% yield. With both building blocks in hand, the rigid rod type target structure \textbf{2} was assembled with a Sonogashira-Hagihara reaction. Thus the central diiodo precursor \textbf{12} and two equivalents of the alkyne \textbf{10} were exposed to catalytic amounts of tetrakis(triphenylphosphine)palladium and copper iodide in a tetrahydrofuran–diisopropylamine mixture. Workup and purification by CC provided \textbf{2} as a white solid in a yield of $22\,$\%.

The target structure \textbf{1} and the precursors \textbf{6}-\textbf{12} were soluble in common organic solvents and were characterized by $^1H$-, $^{13}$C- and $^{19}$F-NMR spectroscopy, elemental analysis (EA)\footnote{EA of \textbf{8} is missing due to limited stability.} and mass spectrometry. $^1$H- and $^{19}$F-NMR spectra of the poorly soluble compound \textbf{2} were recorded in mixtures of deuterated chloroform with fluorinated solvents like hexafluorobenzene or 1,1,2-trichloro-1,2,2-trifluoroethane (Freon 113\textregistered). Furthermore, the identity of \textbf{2} was corroborated by MALDI-ToF mass spectrometry.

\begin{figure}[htbp]
   \centering
   \includegraphics[width=1.0\columnwidth]{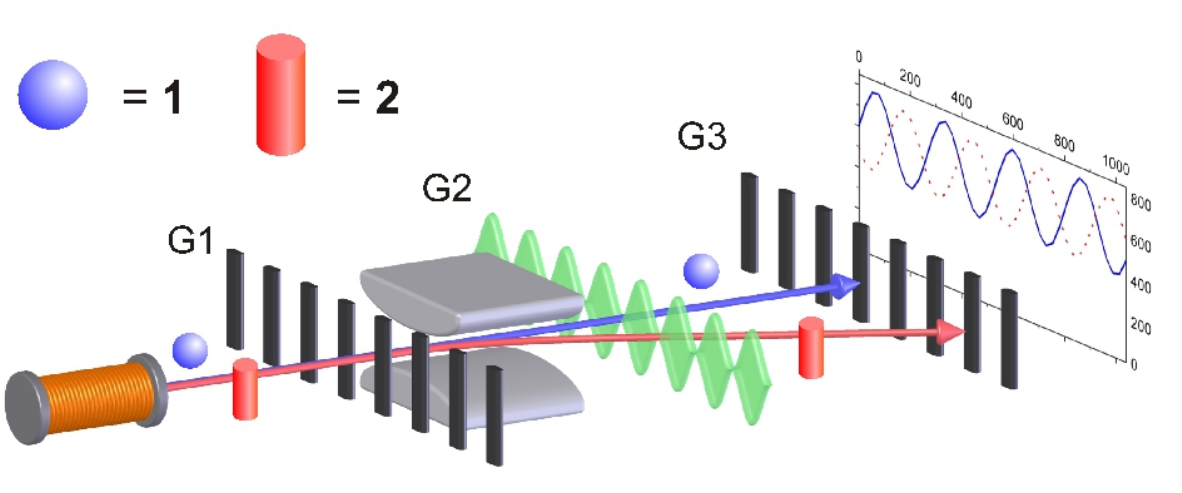}
   \caption{Sketch of the matter wave interferometer that was used to distinguish the constitutional isomers \textbf{1} and \textbf{2}. The isomers were ejected from the effusive source to pass two nano-mechanical gratings and one optical phase grating. Quantum interference leads to a molecular density distribution at G3. Its lateral position is determined by the interaction between the external electric field and the molecular susceptibility $\chi_{tot}$. The susceptibility varies for different atomic arrangements even if they add up to the same chemical sum formula.
} \label{fig:setup}
  \end{figure}

To distinguish between both constitutional isomers, the substances were evaporated under high vacuum conditions in a ceramic furnace at a temperature of $T=185\pm5\,$\textdegree C. The molecular beam passed through a series of delimiters that restricted the trajectory to the free-fall parabola that corresponds to the desired velocity. The particles then traversed a Kapitza-Dirac-Talbot-Lau (KDTL) interferometer~\,\cite{Gerlich2007}, shown in Fig. \ref{fig:setup}, before they were ionized by electron impact and injected into a quadrupole mass spectrometer (QMS).

The matter wave interferometer consists of three gratings, G1-G3, all with identical slit periods of $d=266\,$nm and slit openings as small as about $100\,$nm. The first grating (G1) prepares the necessary lateral wave coherence of the molecular beam at the location of G2. Diffraction of the molecules at G2 then results in a regular density distribution of periodicity d at G3. When G3 is scanned across this molecular pattern, the detector records a sinusoidal intensity variation
\begin{equation}
	S(x_3)=O+A\cdot\sin[2\pi(x_3-\Delta x_3)/d]
	\label{eq:signal}
\end{equation}
as shown at the right-hand end of Fig. \ref{fig:setup}. Here, $x_3$ denotes the grating position, $\Delta x_3$ is the position offset of the interference fringe, which also depends on external forces, and we define the quantum interference visibility $V=A/O$ as the ratio of the fringe amplitude $A$ and its vertical offset $O$.

The three gratings are spaced equidistantly at a distance of $105\,$mm one from another. G1 and G3 are absorptive masks that are fabricated from a $190\,$nm thin SiN$_x$ membrane. The central grating, which is responsible for diffraction, is realized as a standing light wave that is generated by a retro-reflected laser beam at $\lambda\approx532\,$nm. The light field imposes a position dependent phase on the matter wave that is governed by the laser power $P$ and the optical polarizability $\alpha_{opt}$ at the chosen laser wavelength. For many molecules this value approximates very closely the static value $\alpha_{stat}$, when the dipole allowed electronic resonances are sufficiently separated from the laser wavelength. This is also the case for our structures \textbf{1} and \textbf{2}.

The application of a pair of electrodes between gratings G1 and G2 allows us to subject the matter wave to a homogeneous electric force field that is directed along the k-vector of the grating laser and which is constant to within $1\,$\% across the molecular beam. The electric field imprints an additional phase onto the matter wave that results in a shift $\Delta x_3$ of the interference pattern at G3 parallel to the grating vector. This shift is proportional to the total electric susceptibility $\chi_{tot}$:
\begin{equation}
	\Delta x_3\propto\chi_{stat}U^2/(mv^2)
	\label{eq:chi}
\end{equation}
where $m$ is the mass, $v$ is the molecular velocity and $U$ the voltage applied between the electrodes~\,\cite{Heer2010}. In order to obtain absolute numbers, a geometry factor has to be determined experimentally, in our case in a calibration measurement with C$_{60}$.

Generally, the total electric susceptibility is determined by the electronic contribution to the polarizability $\alpha_{stat}$ and, according to the van Vleck formula~\,\cite{Vleck1965}
\begin{equation}
	\chi_{tot}=\alpha_{stat}+\left\langle d^2\right\rangle/(3k_BT)
	\label{eq:chitot}
\end{equation}
also by a thermal average over the square of all possible electric dipole moments, be they permanent moments or those related to thermally activated vibrations~\,\cite{Compagnon2002}.

As the polarizability is influenced by the molecular structure, in our case by the extent of delocalization of the central $\pi$-systems, we expect even constitutional isomers to behave differently under the influence of the external field. Effects of the fluorous ponytails on the polarizability are expected to be negligible mainly for two reasons: First, their electronic coupling to the aromatic subunit is poor, and second their contributions will be comparable for both constitutional isomers \textbf{1} and \textbf{2}.

Compounds \textbf{1} and \textbf{2} were examined in two separate experimental runs. They were evaporated at identical temperatures, but we chose slightly different velocity distributions. These were centered at $v_{mean}=110\,$ms$^{-1}$ for compound \textbf{1} and $v_{mean}=91\,$ms$^{-1}$ for \textbf{2} with $\Delta v_{FWHM}/v_{mean}=0.15$ and $\Delta v_{FWHM}/v_{mean}=0.10$, respectively. This corresponds to a mean de Broglie wavelength of about $2.5\,$pm.

Each individual interference pattern was sampled in steps of $\Delta x_3=26\,$nm over a range of $1064\,$nm, corresponding to four full interference fringe periods. The interference patterns were recorded at different voltage settings.

At high voltages the fringe visibility is reduced by a velocity dependent dephasing of the interference pattern and the finite width of the velocity distribution. Since $\Delta x_3$ is velocity dependent (equation \ref{eq:chi}), a large velocity spread will blur the interference contrast.

In order to minimize the effect of thermal drifts of the grating position, the field-dependent fringe shift was read for each deflection voltage separately. For each position of G3 the molecular beam transmission was measured both at the desired deflection voltage and for a reference value of $U=1\,$kV.

\begin{figure}[htbp]
   \centering
   \includegraphics[width=1.0\columnwidth]{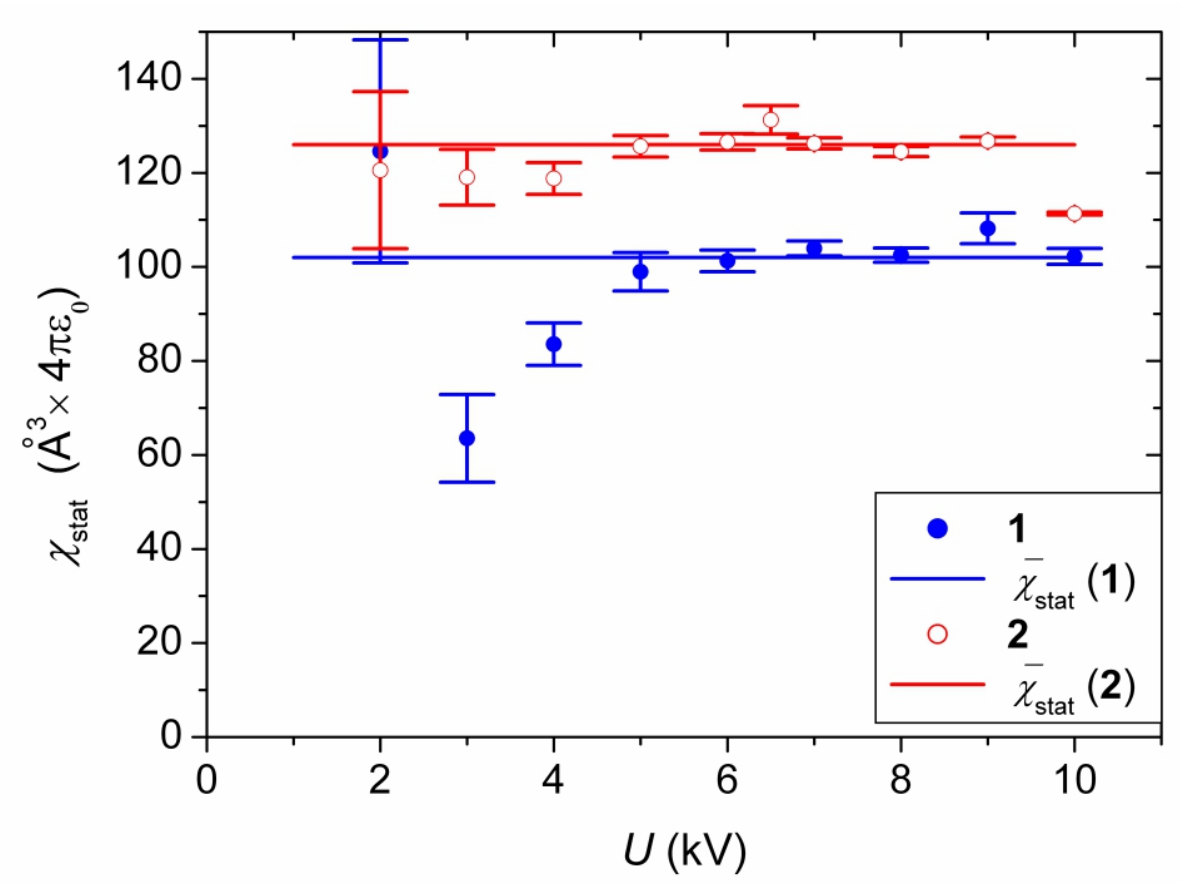}
   \caption{Experimental values of $\chi_{stat}$ for compounds \textbf{1} (blue full circles) and \textbf{2} (red hollow circles) extracted from the interference fringe shift at different settings of the deflection voltage. The error bars represent the statistical errors (1$\sigma$). The solid blue and red lines show the weighted means of the susceptibility values of \textbf{1} and \textbf{2}, respectively \footnote{see Supporting Information at \href{http://dx.doi.org/10.1039/c0cc00125b}{DOI: 10.1039/c0cc00125b}}.
} \label{fig:chi}
  \end{figure}

The electric susceptibility was determined for every single voltage step by fitting equation \ref{eq:chi} to the experimental value of $\Delta x_3$ with $\chi_{stat}$ as the only free parameter. The calculation included the detailed measured velocity distribution. The results of all runs for both molecules are depicted in Fig. \ref{fig:chi}. We find a weighted mean of $\chi_{stat}=102\pm0.8\,\mathrm{\AA}^3\times4\pi\epsilon_0$ for compound \textbf{1} and $\chi_{stat}=126\pm0.5\,\mathrm{\AA}^3\times4\pi\epsilon_0$ for \textbf{2}. We show only the statistical error bar, which decreases because of the more reliable reading at high fringe deflection and high voltages. The systematic error is dominated by the uncertainty in the velocity measurement as well as the knowledge of both laser power and focal width. The drop in interference contrast spoils the fit quality at high deflection voltage. We therefore exclude the data point at $U=10\,$kV in the evaluation of the mean value of $\chi$.

Interestingly, for both isomers the susceptibility values differ from the computed static polarizabilities. This is consistent with the presence of vibration-activated electric dipole moments which emerge for flexible molecules at high temperature~\,\cite{Compagnon2002}. Based on our earlier experiments with perfluoroalkyl-functionalized azobenzenes~\,\cite{Gring2010} we expect a thermal contribution to the susceptibility of $10$-$15\,\mathrm{\AA}^3\times4\pi\epsilon_0$ per side chain.

Summarizing, the different total susceptibilities of both constitutional isomers lead to different de Broglie interference shifts in the presence of external fields. The isomers thus are distinguishable even by pure center-of-mass interferometry, in spite of their identical mass and chemical sum formula.

\begin{acknowledgments}
Our work was supported by the Austrian FWF within the programs Wittgenstein Z149-N16, DK-W1210 CoQuS, the Swiss National Science Foundation and ESF EuroQuasar MIME.
\end{acknowledgments}

%

\end{document}